\documentclass[letterpaper,12pt]{article}
\usepackage{longtable}
\usepackage{lscape}
\usepackage{amsmath}
\usepackage{amssymb}
\usepackage[dvips]{graphicx,epsfig}
\usepackage{psfrag}
\title{Quasinormal modes of $D$-dimensional de Sitter spacetime}

\author{A.\ L\'opez-Ortega\thanks{Electronic address: alfredo@fis.cinvestav.mx} \\Departamento de F\'{\i}sica, CINVESTAV IPN\\ Apdo. Postal 14-740, 07000 M\'exico D. F., M\'exico.  }

\begin{document}

\maketitle

\begin{abstract}

We calculate the exact values of the quasinormal frequencies for an electromagnetic field and a gravitational perturbation moving in $D$-dimensional de Sitter spacetime ($D \geq 4$). We also study the quasinormal modes of a real massive scalar field and compare our results with those of other references.

\end{abstract}

\textbf{Keywords}\,\,\, de Sitter spacetime; Quasinormal modes; Classical fields.

\textbf{PACS numbers} \,\,\, 04.30.-w, 04.30.Nk, 04.40.-b

\textbf{Running title}\,\,\,  de Sitter quasinormal modes


\section{Introduction}

Motivated by Brane World scenario in String Theory and by the study of the higher dimensional features of General Relativity, recently there has been considerable interest in understanding the dynamics of classical fields moving in higher dimensional spacetimes (see Refs.\ \cite{Frolov:2002xf}--\cite{Aros:2002te} for some examples). 

The main reason to study the quasinormal modes (QNMs) of black holes is its possible astrophysical importance; however, in many physical relevant cases an exact analytical computation of the quasinormal (QN) frequencies is not possible and they are calculated by using numerical methods or suitable approximations (see Refs.\ \cite{Kokkotas:1999bd} for reviews). Recently, it was  shown that the QNMs are a useful tool in understanding the AdS-CFT and dS-CFT correspondences and some aspects of quantum gravity \cite{Horowitz:1999jd}, \cite{Motl:2002hd}. For these and other reasons the QN frequencies of several fields have been calculated in some higher dimensional spacetimes \cite{Natario:2004jd}--\cite{Aros:2002te}.

From a theoretical viewpoint, it is useful to have a list of examples in which we can exactly compute the QN frequencies to achieve a better understanding of its properties. In Refs.\ \cite{Natario:2004jd}--\cite{Vanzo:2004fy}, \cite{Molina:2003ff}, \cite{Aros:2002te}, \cite{Brady:1999wd}--\cite{Cardoso:2003sw} we can find some examples of QN frequencies analytically calculated.

As is well known, the usual method to study the physical properties of a spacetime is to investigate how a classical field propagates in it. Owing to the simplicity of its metric \cite{Spradlin:2001pw}, de Sitter background has many applications in Theoretical Physics \cite{Park:1998qk}, \cite{dS3d:inflation-papers}, \cite{Strominger:2001pn}. Even more, in this background the dynamics of the field is simpler than in other physical relevant metrics. For this reason, in Refs.\ \cite{Brady:1999wd}--\cite{Lopez-Ortega:2006ig}, \cite{dS3D-Lohiya}--\cite{Lopez-Ortega:2004cq} the propagation of classical fields in $4D$ de Sitter spacetime was analyzed. Furthermore, in some of these references the QNMs of the de Sitter spacetime were defined and analytically calculated. To our knowledge they were first computed in Ref.\ \cite{Brady:1999wd}, where it is studied the effect of the cosmological horizon on the late time behavior of the scalar field moving in the Schwarzschild de Sitter (SdS) black hole. More recently the dS-CFT correspondence also motivated the calculation of these de Sitter QN frequencies (see \cite{Abdalla:2002hg} and references therein).

Here we study  how change the results on the QN frequencies given in previous references when the dimension of de Sitter spacetime is greater than four. For previous work along this line see Ref.\ \cite{Natario:2004jd} for the computations corresponding to the gravitational perturbations and Refs.\ \cite{Du:2004jt}, \cite{Abdalla:2002hg} for those of the scalar field.

Nevertheless, it is convenient to say that these papers report different conclusions on the existence of the de Sitter QNMs for massless fields. For example, in Ref.\ \cite{Natario:2004jd} Nat\'ario and Schiappa asserted that for a gravitational perturbation they are well defined only in odd spacetime dimensions; in \cite{Brady:1999wd} Brady \textit{et al.\ }claimed that for a minimally coupled massless Klein-Gordon field there are no well defined QN frequencies in four dimensions;  in Ref.\ \cite{Du:2004jt} Du, \textit{et al.\ }found that a massive Klein-Gordon field have well defined QNMs only when its mass satisfies a condition and in Ref.\ \cite{Myung:2003ki} Myung and Kim claimed that these QNMs do not exist. 

In Ref.\ \cite{Lopez-Ortega:2006ig} we show that in four dimensions there are well defined de Sitter QNMs for the massless fields of spin $\tfrac{1}{2}$, $1$, and $2$; but this conclusion disagree with the affirmations of \cite{Natario:2004jd} (see also Ref.\ \cite{Du:2004jt}).\footnote{The de Sitter QN frequencies for the Dirac field in three and four dimensions are computed in Refs.\ \cite{Du:2004jt}, \cite{Lopez-Ortega:2006ig}.} Motivated by this discussion, in the present paper we calculate these QNMs for an electromagnetic field, a gravitational perturbation and a massive scalar field in $D \geq 4$ dimensions. Thus we extend to $D$-dimensional de Sitter spacetime some results of Ref.\ \cite{Lopez-Ortega:2006ig} and make some comments on the issues enumerated above. 

We organize this paper as follows. In Secs.\ \ref{s:electromagnetic-field}, \ref{s:gravitational perturbations}, and \ref{s: Klein Gordon} for an electromagnetic field, a gravitational perturbation, and a real Klein-Gordon field moving in $D$-dimensional de Sitter background we find the exact values of its QN frequencies. In Section \ref{s: Discussion} we discuss the results found in the previous sections. 

In Appendix \ref{dsd:appendix-mathematical} we enumerate some properties of the hypergeometric function that we use in several sections of this paper. In the next two Appendices we discuss some facts slightly outside the main line of the paper. In Appendix \ref{app: near extreme quasinormal}, following Ref.\ \cite{Molina:2003ff}, we calculate the QN frequencies of an electromagnetic field and a gravitational perturbation propagating in $D$-dimensional near extreme SdS black hole. Finally, in Appendix \ref{dsd:app-effective-potential} we discuss some facts on the effective potentials of the Schr\"odinger type radial equations in which an analytical calculation of the QN frequencies is possible.

\section{Electromagnetic field}
\label{s:electromagnetic-field}

In the following we study the behavior of different classical fields moving in $D$-dimensional de Sitter spacetime. We use the metric of the de Sitter background in static coordinates
\begin{equation} \label{e:metric-general}
{\rm d}s^2=P(r)^2 {\rm d}t^2 - \frac{{\rm d} r^2}{P(r)^2} -r^2 {\rm d} \Omega^2,
\end{equation}
where ${\rm d} \Omega^2 = \gamma_{ij} {\rm d}x^i {\rm d} x^j$ is the line element of the $(D-2)$-dimensional unit sphere $(i,j,\ldots =1,2,\dots,D-2)$ and
\begin{equation} \label{e:definition P}
P^2 = 1 -\frac{r^2}{L^2}.
\end{equation} 

Using a modified Feynman gauge, in Ref.\ \cite{Crispino:2000jx} Crispino, \textit{et al.}, showed that the equations of motion for an electromagnetic field propagating in a spherically symmetric and static spacetime (that is, for a background of the form (\ref{e:metric-general}) where the quantity $P^2 $ is not necessarily given by (\ref{e:definition P})) reduce to a decoupled set of radial differential equations. Here we study the so called physical modes \textbf{I} and \textbf{II} which represent the physical degrees of freedom of an electromagnetic field. For more details see Ref.\ \cite{Crispino:2000jx}.

\subsection{Physical modes \textbf{I}}
\label{s:modes-I}

For these modes, the vector potential $A_\alpha$ in the metric (\ref{e:metric-general}) is equal to \cite{Crispino:2000jx}
\begin{align} \label{e:modes-I-vectorial}
A_t^{(I)} &= 0, \nonumber \\
A_r^{(I)}  &= R^{(I)}(r)\, Y_{lm}\, \textrm{e}^{-i \omega t}, \\
A_i^{(I)}  &= \frac{r^{4-D}}{l(l+D-3)} P^2 \frac{{\rm d}}{{\rm d} r}[r^{D-2}R^{(I)}]\, \partial_i Y_{lm} \, \textrm{e}^{-i\omega t}, \nonumber 
\end{align} 
where $Y_{lm}$ stands for the scalar spherical harmonics on the $(D-2)$-di\-men\-sion\-al unit sphere \cite{Crispino:2000jx}, \cite{Chodos:1983zi}, and the radial function $R^{(I)}$ satisfies (Eq.\ (2.18) of Ref.\ \cite{Crispino:2000jx})
\begin{eqnarray} \label{e:radial-I}
\frac{1}{r^2}\frac{{\rm d}}{{\rm d} r}\left[P^2 r^{4-D} \frac{{\rm d}}{{\rm d} r}\left( r^{D-2} R^{(I)} \right) \right] + \left[\frac{\omega^2}{P^2}-\frac{l(l+D-3)}{r^2} \right] R^{(I)} = 0.
\end{eqnarray} 
In order to satisfy the gauge conditions $l$ must satisfy $l \geq 1$ \cite{Crispino:2000jx}.

The previous equation in $D$-dimensional de Sitter metric takes the form
\begin{align} \label{e:modes-I-electromagnetic}
(1-z^2)&\frac{{\rm d}^2 R^{(I)}}{{\rm d}z^2} + \left[\frac{D}{z}-(D+2)z \right] \frac{{\rm d} R^{(I)}}{{\rm d}z} \nonumber \\ 
&+\left[ \frac{\tilde{\omega}^2}{1-z^2} - \frac{(l-1)(l+D-2)}{z^2} -3(D-2)\right] R^{(I)} = 0,
\end{align} 
where $z=r/L$ and $\tilde{\omega}=\omega L$. Making in Eq.\ (\ref{e:modes-I-electromagnetic}) the change of variable $y=z^2$ \cite{Lopez-Ortega:2006ig}, it is transformed into
\begin{align}
\frac{{\rm d}^2 R^{(I)}}{{\rm d}y^2} &+\left[ \frac{(D+1)}{2y} -\frac{1}{1-y} \right] \frac{{\rm d} R^{(I)}}{{\rm d}y} \nonumber \\
&+\frac{1}{4}\left[\frac{\tilde{\omega}^2}{y(1-y)^2} -\frac{(l-1)(l+D-2)}{y^2 (1-y)} - \frac{3(D-2)}{y(1-y)} \right] R^{(I)} = 0.
\end{align}

For the function $R^{(I)}$ we make the ansatz\footnote{The symbols $A$, $B$, $a$, $b$, $c$, and $\tilde{R}$ represent different quantities in each section of this paper.}
\begin{equation} \label{e: radial ansatz}
R^{(I)} = y^A (1-y)^B \tilde{R},
\end{equation} 
where
\begin{eqnarray}
A &= \left\{ \begin{array}{l} \frac{l-1}{2} , \\  \\ - \frac{l+D-2}{2}, \end{array}\right. \qquad \qquad B= \pm \frac{i \tilde{\omega}}{2},
\end{eqnarray} 
to find that the function $\tilde{R}$ must be a solution to the hypergeometric differential equation
\begin{equation} \label{e:hypergeometric-modes-I}
y(1-y) \frac{{\rm d}^2 \tilde{R}}{{\rm d}y^2} + (c - (a +b + 1)y)\frac{{\rm d} \tilde{R}}{{\rm d} y} - a b \tilde{R}   = 0,
\end{equation} 
where the parameters $a$, $b$, and $c$ are equal to
\begin{align} \label{e:a-b-c-values-modes-I}
a &= A + B + \frac{D-2}{2} ,\nonumber \\
b &= A + B + \frac{3}{2}, \\
c &= 2 A + \frac{D+1}{2}. \nonumber
\end{align}

As noticed in Refs.\ \cite{Abdalla:2002hg}, \cite{Lopez-Ortega:2006ig}, \cite{Lopez-Ortega:2004cq} there are several possible combinations for the values of $A$ and $B$ (and therefore of $a$, $b$, and $c$). Thus the solutions of Eq.\ (\ref{e:hypergeometric-modes-I}) can be written in several equivalent forms. In the following we study in detail the case $A= \tfrac{l-1}{2}$ and $B=\tfrac{i \tilde{\omega}}{2}$.

The first solution to Eq.\ (\ref{e:hypergeometric-modes-I}) is given by (see Appendix \ref{dsd:appendix-mathematical} for notation)
\begin{equation}
\tilde{R}_1 = {}_{2}F_{1}(a,b;c;y).
\end{equation} 
If the parameter $c$ is a half-integer (even $D$), then the second solution to Eq.\ (\ref{e:hypergeometric-modes-I}) is 
\begin{equation}
\tilde{R}_2 = y^{1-c} {}_{2}F_{1}(a-c+1,b-c+1;2-c;y),
\end{equation} 
and if $c$ is an integer ($D$ odd), then the second solution to Eq.\ (\ref{e:hypergeometric-modes-I}) is equal to
\begin{align} 
\tilde{R}_2 &= {}_{2}F_{1}(a,b;c;y) \ln (y) \nonumber \\
& + \frac{(c - 1)!}{\Gamma(a) \Gamma(b)} \sum_{s=1}^{c - 1} (-1)^{s-1} (s - 1)! \frac{\Gamma(a - s) \Gamma(b -s)}{(c -s - 1)!} y^{-s} \nonumber \\
&  + \sum_{s=0}^{\infty} \frac{(a)_s (b)_s}{s! (c)_s} y^s \left[\psi(a + s) + \psi(b + s) - \psi(c + s) - \psi(1 + s) \right. \nonumber \\
  & \left. \hspace{0cm} - \psi(a -  c +1) - \psi(b - c + 1) + \psi(1) + \psi(c -1) \right]. 
\end{align} 

We have no reasons to expect singularities of the field at $r=0$; therefore in this work we use radial functions that are regular there. For even and odd $D$, the only solution to Eq.\ (\ref{e:hypergeometric-modes-I}) that leads to a regular radial function at $r=0$ is $\tilde{R}_1$ \cite{Lopez-Ortega:2006ig}, \cite{Lopez-Ortega:2004cq}. The radial function that includes $\tilde{R}_2$ is divergent at $r=0$. Thus the regular radial function at the origin takes the form
\begin{equation} \label{e:physical-solution-modes-I}
R_1^{(I)} = y^{\tfrac{l-1}{2}} (1-y)^{\tfrac{i \tilde{\omega}}{2}} {}_{2}F_{1}(a,b;c;y).
\end{equation} 

As in Refs.\ \cite{Natario:2004jd}, \cite{Du:2004jt}, \cite{Brady:1999wd}--\cite{Lopez-Ortega:2006ig} the de Sitter QNMs are solutions to the equations of motion that satisfy the physical boundary conditions: (i) the field is regular at the origin; (ii) the field is purely outgoing near the cosmological horizon. To calculate the QN frequencies of the electromagnetic field modes \textbf{I} we exploit Eq.\ (\ref{e:hypergeometric-property-z-1-z}) of Appendix \ref{dsd:appendix-mathematical}, which is valid when the quantity $c-a-b$ is not an integer, to expand the radial function (\ref{e:physical-solution-modes-I}) in the form
\begin{align} \label{e:physical-field-1-y}
R_1^{(I)} &= y^{\tfrac{l-1}{2}}\left\{\frac{\Gamma(c)\Gamma(c-a-b)}{\Gamma(c-a)\Gamma(c-b)}(1-y)^{\tfrac{i \tilde{\omega}}{2}} {}_{2}F_{1}(a,b;a+b-c+1;1-y)\right. \nonumber\\ 
&+ \left.\frac{\Gamma(c)\Gamma(a+b-c)}{\Gamma(a)\Gamma(b)}(1-y)^{-\tfrac{i \tilde{\omega}}{2}} {}_{2}F_{1}(c-a,c-b;c-a-b+1;1-y) \right\}.  
\end{align} 

Using Eq.\ (\ref{e:modes-I-vectorial}), we find that the first term in curly brackets of Eq.\ (\ref{e:physical-field-1-y}) represents an ingoing wave and the second term represents an outgoing wave. To satisfy the boundary conditions of the de Sitter QNMs we must impose the condition  \cite{Natario:2004jd}, \cite{Brady:1999wd}, \cite{Choudhury:2003wd}
\begin{equation} \label{e:c-a-condition-modes-I}
c-a= -n, \quad \qquad \textrm{or} \quad \qquad c-b=-n, \quad \qquad n=0,1,2,\dots
\end{equation} 
But from the values of the parameters $a$, $b$, and $c$ given in Eq.\ (\ref{e:a-b-c-values-modes-I}), if the quantity $c-a-b = -i \tilde{\omega}$ is not an integer (to expand (\ref{e:physical-solution-modes-I}) as in (\ref{e:physical-field-1-y})), then the quantities $c-a = (l+2-i \tilde{\omega})/2$ and $c-b=(l+D-3-i \tilde{\omega})/2$ cannot be integers, since $(l+2)/2$ and $(l+D-3)/2$ are integers or half-integers, $-i \tilde{\omega}/2$ is not an integer or half-integer and therefore their sum cannot be an integer. Hence we cannot satisfy the conditions (\ref{e:c-a-condition-modes-I}) when $c-a-b$ is not an integer. 

Notice that if we impose either of the conditions $c-a=-n$, or $c-b=-n$, then we contradict the assumption that $c-a-b$ is not an integer. To write the radial function $R_1^{(I)}$ in the form (\ref{e:physical-field-1-y}) we need to make this assumption in order to use Eq.\ (\ref{e:hypergeometric-property-z-1-z}) of Appendix \ref{dsd:appendix-mathematical}.

This also happens when we calculate the de Sitter QN frequencies of massless fields in four dimensions as shown in Section 3.2 of \cite{Lopez-Ortega:2006ig}. In that work for the massless fields of spin $\tfrac{1}{2}$, $1$, and $2$ moving in four dimensional de Sitter spacetime was noted that if $c^+-a^+-b^+=-(i \tilde{\omega} -|s|)$ is not an integer, then $c^+-a^+=j+1-|s|-(i \tilde{\omega} -|s|)$ and $c^+-b^+=j+1+|s|$ cannot be negative integers.\footnote{In this paragraph we use the notation of \cite{Lopez-Ortega:2006ig}.} (See also Appendix B of \cite{Lopez-Ortega:2006ig}.) Furthermore, in this reference were computed the de Sitter QN frequencies of a massless Dirac field in three dimensions and a massive Dirac field in four dimensions. Notice that for the massless Dirac field in three dimensions we can fulfill the QNMs boundary conditions  when the quantity $c-a-b$ is not an integer (see Section 2.1 of \cite{Lopez-Ortega:2006ig}).

Therefore, as in Section 3.2 of Ref.\ \cite{Lopez-Ortega:2006ig}, we must study the case when $c-a-b$ is an integer. If $c-a-b=-n_1$, where $n_1=1,2,3,\dots$, we can write the radial function (\ref{e:physical-solution-modes-I}) as\footnote{For $c-a-b$ equal to zero or a positive integer, the field is static or its amplitude increases with time \cite{Lopez-Ortega:2006ig}.} (see Appendix \ref{dsd:appendix-mathematical})
\begin{align} \label{e:hypergeometric-integer-I}
&R_1^{(I)}  =  y^{\tfrac{l-1}{2}} (1-y)^{\tfrac{i \tilde{\omega}}{2}} \left\{ \frac{\Gamma(a+b-n_1) \Gamma(n_1)}{\Gamma(a)\Gamma(b)} (1-y)^{-n_1} \right.  \\ 
& \times \sum_{s=0}^{n_1-1}\frac{(a-n_1)_s (b-n_1)_s}{s! (1-n_1)_s}(1-y)^s  
- \frac{(-1)^{n_1} \Gamma(a+b-n_1)}{\Gamma(a-n_1)\Gamma(b-n_1)} \sum_{s=0}^\infty \frac{(a)_s (b)_s}{s!(n+s)!}   \nonumber \\
&\left. \times (1-y)^s [\textrm{ln}(1-y) -\psi(s+1) -\psi(s+n+1)+\psi(a+s)+\psi(b+s)] \right\} .\nonumber
\end{align} 

From Eq.\ (\ref{e:modes-I-vectorial}), we find that the second term in curly brackets of (\ref{e:hypergeometric-integer-I}) represents an ingoing wave while the first term represents an outgoing wave. Thus to satisfy the boundary conditions of the de Sitter QNMs we must impose the condition
\begin{equation} \label{e:conditions-quasinormal-I}
a- n_1 = -n, \quad \qquad \textrm{or} \quad \qquad b - n_1 = -n.
\end{equation} 
The previous conditions imply that the de Sitter QN frequencies are equal to\footnote{In this paper, we loosely call QN frequencies to the quantities $i\tilde{\omega}$. We can easily calculate $\omega$ from $i\tilde{\omega}$.}
\begin{equation} \label{e:quasinormal-frequency-I}
i\tilde{\omega} = l + D-3 + 2 n, \qquad \qquad i\tilde{\omega} = l + 2 + 2 n.
\end{equation} 
Notice that for these QN frequencies we do not contradict our initial assumptions.

In the case studied in Ref.\ \cite{Abdalla:2002hg} was shown that identical de Sitter QN frequencies are found when they use different values for the parameters $A$, $B$, $a$, $b$, and $c$. We believe that this also happens for the classical fields that are analyzed here \cite{Lopez-Ortega:2006ig}.

\subsection{Physical modes \textbf{II}}
\label{s:modes-II}

For these modes the vector potential in the metric (\ref{e:metric-general}) is equal to \cite{Crispino:2000jx}
\begin{align}
A_t^{(II)} &= A_r^{(II)} = 0, \nonumber \\
A_i^{(II)} &= R^{(II)}(r)\, Y_i^{lm}\, \textrm{e}^{-i \omega t},
\end{align} 
where $Y_i^{lm}$ stands for the divergence-free vector spherical harmonics on the unit $(D-2)$-sphere \cite{Crispino:2000jx}, \cite{Chodos:1983zi}, \cite{b:Camporesi-harmonics} and the radial function $R^{(II)}$ is a solution to the differential equation (Eq.\ (2.21) of Ref.\ \cite{Crispino:2000jx})
\begin{equation} \label{e:radial-II}
\frac{1}{r^{D-4}} \frac{{\rm d}}{{\rm d} r} \left[P^2 r^{D-4} \frac{{\rm d}}{{\rm d} r}\right] R^{(II)} + \left[ \frac{\omega^2}{P^2}- \frac{(l+1)(l+D-4)}{r^2}\right] R^{(II)} = 0 .
\end{equation} 

In $D$-dimensional de Sitter metric we can solve this equation and using a similar procedure to that of the previous subsection we compute the QN frequencies. Therefore we only present the main steps. Employing the variable $y$ previously defined, it is possible to find exact solutions to Eq.\ (\ref{e:radial-II}) of the form (\ref{e: radial ansatz}), when the quantities $A$ and $B$ are
\begin{eqnarray}
A &= \left\{ \begin{array}{l} \frac{l+1}{2} , \\  \\ - \frac{l+D-4}{2}, \end{array}\right. \qquad \qquad B= \pm \frac{i \tilde{\omega}}{2},
\end{eqnarray} 
and the function $\tilde{R}$ is now a solution to the hypergeometric differential equation (\ref{e:hypergeometric-differential}) with the parameters $a$, $b$, and $c$ equal to
\begin{align} \label{e:a-b-c-values-modes-II}
a &= A + B + \frac{D-3}{2} ,\nonumber \\
b &= A + B, \\
c &= 2 A + \frac{D-3}{2}. \nonumber
\end{align}

Again, we have several possible combinations for the values of parameters $A$ and $B$. In this subsection we study in detail the case $A=\tfrac{l+1}{2}$ and $B=\tfrac{i \tilde{\omega}}{2}$. As in previous subsection we can show that the regular radial function at $r=0$ is
\begin{equation} \label{e:physical-solution-modes-II}
R^{(II)} = y^{\tfrac{l+1}{2}} (1-y)^{\tfrac{i \tilde{\omega}}{2}} {}_{2}F_{1}(a,b;c;y).
\end{equation} 
The solutions to the differential equation for $\tilde{R}$ of the form (\ref{e:solution-two-I}) ($c$ a half-integer, even $D$) or (\ref{e:solution-two-II}) ($c$ an integer, odd $D$) lead to an irregular radial function at $r=0$. 

From the values for $a$, $b$, and $c$ given in Eq.\ (\ref{e:a-b-c-values-modes-II}), if the quantity $c-a-b$ is not an integer, then we cannot satisfy the de Sitter QNMs boundary conditions since the quantities $c-a$ and $c-b$ cannot be integers; thus we must calculate the QN frequencies when $c-a-b$ is a negative integer (see previous subsection and Section 3.2 of Ref.\ \cite{Lopez-Ortega:2006ig}). For this case, applying a similar procedure to that of Subsection \ref{s:modes-I}, we expand the radial function $R^{(II)}$ as in Eq.\ (\ref{e:hypergeometric-integer-I}), and then impose the conditions (\ref{e:conditions-quasinormal-I}) to fulfill the de Sitter QNMs boundary conditions. Therefore we find that the QN frequencies for an electromagnetic mode of type \textbf{II} are equal to
\begin{equation} \label{e:quasinormal-frequency-II}
i\tilde{\omega} = l + D-2 + 2 n, \qquad \qquad i\tilde{\omega} = l + 1 + 2 n.
\end{equation} 

According to the results of Ref.\ \cite{Natario:2004jd}, the de Sitter QN frequencies for a gravitational perturbation are well defined only when the spacetime dimension $D$ is odd (see Section \ref{s:gravitational perturbations} below). Our results (\ref{e:quasinormal-frequency-I}) and (\ref{e:quasinormal-frequency-II}) show that for an electromagnetic field its QN frequencies are well defined in even and odd dimensions. We also note that the QN frequencies of the electromagnetic field have not been previously calculated.

The QN frequencies (\ref{e:quasinormal-frequency-I}) and (\ref{e:quasinormal-frequency-II}) for the modes \textbf{I} and \textbf{II} of an electromagnetic field are equal when $D=4$. For even $D$ (and for given $l$) the QN frequencies (\ref{e:quasinormal-frequency-I}) and (\ref{e:quasinormal-frequency-II}) are equal except for some frequencies whose number depends on the spacetime dimension. For odd $D$, the de Sitter QN frequencies of the modes \textbf{I} and \textbf{II} have different parity for given $l$, hence these are not equal.

To finish the present section, we note that in metric (\ref{e:metric-general}) the equations (\ref{e:radial-I}) and (\ref{e:radial-II}) can take the form \cite{Crispino:2000jx} 
\begin{equation} \label{e:Schrodinger-equation-I-II}
\left\{ \frac{{\rm d}^2 }{{\rm d} r_*^2}  + \omega^2 - V^{(I,II)}\right\} \Phi^{(I,II)} = 0,
\end{equation} 
where $r_*$ stands for the tortoise coordinate 
\begin{equation} \label{e:tortoise-coordinate}
r_* = \int \frac{{\rm d} r}{P^2},
\end{equation} 
the relation between $\Phi^{(I)}$ and $R^{(I)}$ ($\Phi^{(II)}$ and $R^{(II)}$) is \cite{Crispino:2000jx}
\begin{align}
R^{(I)} &= \frac{\Phi^{(I)}}{r^{D/2}}, \nonumber \\
R^{(II)} &= \frac{\Phi^{(II)}}{r^{(D-4)/2}}, 
\end{align} 
and the effective potentials $V^{(I,II)}$ are equal to \cite{Crispino:2000jx}
\begin{align} \label{e:effective-potentials-I-II}
V^{(I)} &= \frac{l(l+D-3)P^2}{r^2} + \frac{(D-2)(D-4)}{4r^2}P^4 - \frac{(D-4)P^2}{2r}\frac{{\rm d}P^2}{{\rm d}r} , \\
V^{(II)} &= \frac{(l+1)(l+D-4)P^2}{r^2} + \frac{(D-4)(D-6)}{4r^2}P^4 + \frac{(D-4)P^2}{2r}\frac{{\rm d}P^2}{{\rm d}r}. \nonumber
\end{align} 
In de Sitter metric the expressions for the effective potentials (\ref{e:effective-potentials-I-II}) reduce to (see also Refs.\ \cite{Natario:2004jd}, \cite{Du:2004jt}, \cite{Lopez-Ortega:2006ig})
\begin{align} \label{e:Sitter-potentials}
V^{(I)} &= \frac{1}{L^2}\left\{\frac{l(l+D-3) + \tfrac{(D-2)(D-4)}{4}}{\sinh^2(r_*/L)} - \frac{\tfrac{(D-4)(D-6)}{4}}{\cosh^2(r_*/L)} \right\}, \\
V^{(II)} &= \frac{1}{L^2}\left\{\frac{(l+1)(l+D-4) + \tfrac{(D-4)(D-6)}{4}}{\sinh^2(r_*/L)} - \frac{\tfrac{(D-2)(D-4)}{4}}{\cosh^2(r_*/L)} \right\}. \nonumber
\end{align}

For $D=4$  the de Sitter QN frequencies (\ref{e:quasinormal-frequency-I}) and (\ref{e:quasinormal-frequency-II}) are equal to those calculated for an electromagnetic field in Appendix B of Ref.\ \cite{Lopez-Ortega:2006ig}. We note that for $D=4$ the effective potentials (\ref{e:Sitter-potentials}) are equal to that given in the previous reference for the same field.

\section{Gravitational perturbations}
\label{s:gravitational perturbations}

By applying a gauge invariant formalism, in Refs.\ \cite{Kodama:2000fa}--\cite{Kodama:2003kk} Kodama and Ishibashi showed that in four or more spacetime dimensions, the equations of motion for the gravitational perturbations of a maximally symmetric vacuum black hole reduce to Schr\"odinger type equations, (one equation for each perturbation type, see also \cite{Gibbons:2002pq}).\footnote{We note that in Refs.\ \cite{Kodama:2000fa}--\cite{Kodama:2003kk} the signature used is ``mostly plus'', that is, $(-+++)$.} We easily see that these equations are valid in $D$-dimensional de Sitter spacetime as noticed in \cite{Natario:2004jd}. Moreover, in this reference Nat\'ario and Schiappa showed that in this background the radial equations for a scalar type, a vector type and a tensor type gravitational perturbation take the form (see Section 4.3 of Ref.\ \cite{Natario:2004jd})
\begin{equation} \label{e:radial-gravitational}
(1-z^2)\frac{{\rm d}}{{\rm d}z}\left[(1-z^2)\frac{{\rm d}R_G}{{\rm d}z}\right] + \left(\tilde{\omega}^2 - \frac{(1-z^2)}{z^2}\beta(\beta+1)+(1-z^2)\alpha\right)R_G=0,
\end{equation} 
where
\begin{align} \label{e:alpha value}
\alpha = \left\{ \begin{array}{l l} \frac{(D-2)D}{4} ,  & \textrm{tensor type}, \\ \\ \frac{(D-4)(D-2)}{4}, & \textrm{vector type}, \\ \\ \frac{(D-6)(D-4)}{4}, & \textrm{scalar type}, \end{array} \right. 
\end{align}
and
\begin{equation} \label{e:beta value}
\beta= \frac{2l + D - 4}{2}.
\end{equation} 

Using the exact solutions to these radial equations, in Ref.\ \cite{Natario:2004jd} Nat\'ario and Schiappa calculated the de Sitter QN frequencies for a gravitational perturbation. Here we again compute them and comment on the differences between our results and those of the previous reference.

As in previous subsections, it is possible to find exact solutions to the radial differential equation (\ref{e:radial-gravitational}) of the form (\ref{e: radial ansatz}) \cite{Natario:2004jd}, where the parameters $A$ and $B$ are equal to
\begin{eqnarray}
A &= \left\{ \begin{array}{l} \frac{\beta + 1}{2} , \\ \\ - \frac{\beta}{2}, \end{array}\right. \qquad \qquad B= \pm \frac{i \tilde{\omega}}{2},
\end{eqnarray}
and the function $\tilde{R}$ is a solution to the hypergeometric differential equation (\ref{e:hypergeometric-differential}) with the quantities $a$, $b$, and $c$ given by \cite{Natario:2004jd}
\begin{align} \label{e:a-b-c-values-gravitational}
a &= A + B +\frac{1}{4} + \frac{1}{4}\sqrt{1+4\alpha} ,\nonumber \\
b &= A + B +\frac{1}{4} - \frac{1}{4}\sqrt{1+4\alpha} , \\
c &= 2 A + \frac{1}{2}. \nonumber
\end{align}

Depending on the values chosen for $A$ and $B$ the parameters $a$, $b$, and $c$ (and therefore the solutions to radial equations) can take different forms. In the following we study in detail the case $A=\tfrac{\beta +1}{2}$ and $B=\tfrac{i \tilde{\omega}}{2}$. So the parameter $c$ is a positive integer (for odd $D$) or a positive half integer (for even $D$). Therefore the hypergeometric type equation for the function $\tilde{R}$ has a first solution of the form (\ref{e:first-solution}), and depending upon the value of the quantity $c$, the second solution is given by (\ref{e:solution-two-I}) (for even $D$) or is of the form (\ref{e:solution-two-II}) (for odd $D$).\footnote{This fact was not noted in Ref.\ \cite{Natario:2004jd} where for even and odd $D$ the second solution is taken equal to expression (\ref{e:solution-two-I}).} But as in Section \ref{s:electromagnetic-field} we find that the only regular radial function at $r=0$ is
\begin{equation} \label{e:solution-gravitational}
R_G = y^{\tfrac{\beta+1}{2}} (1-y)^{\tfrac{i \tilde{\omega}}{2}} {}_{2}F_{1}(a,b;c;y),
\end{equation} 
because the radial functions that include solutions to the hypergeometric equation of the form (\ref{e:solution-two-I}) or (\ref{e:solution-two-II}) have divergences at the origin of the de Sitter metric (see previous section and Ref.\ \cite{Lopez-Ortega:2006ig}).

When the quantity $c-a-b$ is not integer, according to Eq.\ (\ref{e:hypergeometric-property-z-1-z}) to find a purely outgoing wave as $r\to L$ we must impose the condition $c-a=-n$ or $c-b=-n$, but from the values of $a$, $b$, and $c$ given in Eqs.\ (\ref{e:a-b-c-values-gravitational}), it is not possible to satisfy the previous conditions when $c-a-b$ is not an integer (see Section \ref{s:electromagnetic-field} and paragraphs following to Eq.\ (\ref{e: values q}) below). Therefore, as in Subsection \ref{s:modes-I}, if $c-a-b$ is a negative integer we can use Eq.\ (\ref{e:hypergeometric-property-integer}) to expand the solution (\ref{e:solution-gravitational}) as in formula (\ref{e:hypergeometric-integer-I}). To fulfill the de Sitter QNMs boundary conditions we need to impose the conditions (\ref{e:conditions-quasinormal-I}) and from them we get that the QN frequencies for the gravitational perturbations are equal to
\begin{equation} \label{e:quasinormal-gravitational}
i\tilde{\omega} = l + D-1-q + 2 n, \qquad \qquad i\tilde{\omega} = l + q + 2 n,
\end{equation} 
where the quantity $q$ takes the following values
\begin{align} \label{e: values q}
q = \left\{ \begin{array}{l l} 0,  & \textrm{tensor type}, \\ 1, & \textrm{vector type}, \\ 2, & \textrm{scalar type}. \end{array} \right. 
\end{align}  
We note that this result on the de Sitter QN frequencies is different from that of Ref.\ \cite{Natario:2004jd}, where well defined QN frequencies for de Sitter spacetime were found only when its dimension $D$ is odd. 

We believe that in Ref.\ \cite{Natario:2004jd} was not noticed the following. (In this paragraph we use the notation of \cite{Natario:2004jd} except for $\alpha$, $\beta$, and $\gamma$ that we denote with the symbols $a$, $b$, and $c$, respectively.) To write the radial function $\Phi(r)$ of Ref.\ \cite{Natario:2004jd} as our Eq.\ (\ref{e:physical-field-1-y}) (see equations previous to Eq.\ (4.7) of \cite{Natario:2004jd}), it is necessary to assume that $c-a-b= -i \tilde{\omega}$ is not integer (see Appendix \ref{dsd:appendix-mathematical} and Chapter 4 of the second reference in \cite{b:DE-books}). Taking into account that the quantity $j$ is an odd integer and defining $k$ by $k=(j+1)/2$ we find that the quantity $c-a=(d+l-k)/2-i \tilde{\omega}/2$ cannot be an integer; since $d$, $k$, and $l$ are integers, $(d+l-k)/2$ is an integer or half-integer, and as we supposed that $-i \tilde{\omega}$ is not an integer, the quantity $-i \tilde{\omega}/2$ is not an integer or half-integer. Therefore their sum in $c-a$ cannot be an integer. 

Thus the condition $c-a=-n$, for $n$ an integer, cannot be fulfilled when the quantity $c-a-b$ is not an integer. This also happens with the condition $c-b=-n$. Nevertheless, in \cite{Natario:2004jd} both conditions were imposed and in this reference was not noted that these conditions contradict its first assumption ($c-a-b$ is not an integer). In \cite{Brady:1999wd}, \cite{Choudhury:2003wd} for the massless, minimally coupled to curvature Klein-Gordon field moving in four dimensional de Sitter spacetime are also imposed these contradictory conditions. As a consequence in previous references is asserted that in four dimensions for this field there are no de Sitter QN frequencies.

Hence our results show that the conclusion of Ref.\ \cite{Natario:2004jd} on the de Sitter QNMs in even dimensional spacetimes is not valid. Nevertheless when the spacetime dimension is odd, the second expression in Eq.\ (\ref{e:quasinormal-gravitational}) is equal to that calculated in \cite{Natario:2004jd}, while the first formula in (\ref{e:quasinormal-gravitational}), except for some frequencies whose number depend upon the spacetime dimension, also produces the frequencies computed in Ref.\ \cite{Natario:2004jd}. The first expression in Eq.\ (\ref{e:quasinormal-gravitational}) was not found in the previous reference. 

In four spacetime dimensions the QN frequencies (\ref{e:quasinormal-gravitational}) for a vector type perturbation and a scalar type perturbation are equal. For even $D$ (and for given $l$) the three types of gravitational perturbations have equal de Sitter QN frequencies, except for a set of frequencies whose number depends on the spacetime dimension. For given $l$ and odd $D$, the parity of the QN frequencies for a vector type perturbation is different from the parity of those corresponding to the scalar type and tensor type perturbations, hence  these sets of frequencies are not equal (see Ref.\ \cite{Natario:2004jd}).
 
Applying the monodromy method \cite{Motl:2002hd}, in Section 3.2 of Ref.\ \cite{Natario:2004jd} Nat\'ario and Schiappa calculated the asymptotic value of the QN frequencies for a gravitational perturbation propagating in SdS black hole. The discrete increments in the inverse relaxation time (gap) calculated there in de Sitter limit are equal to $1/L$. In Ref.\ \cite{Natario:2004jd} is also asserted that the gap for de Sitter QN frequencies is equal to $2/L$ which does not coincide with the value found taking the de Sitter limit of the result for the SdS black hole. The value that we obtain for this quantity is equal to $1/L$ which coincides with that found taking the de Sitter limit of the result for the SdS black hole when $n$ goes to infinity.

We note that the radial differential equation (\ref{e:radial-gravitational}) can take the form (\ref{e:Schrodinger-equation-I-II}), with effective potential equal to \cite{Natario:2004jd}
\begin{equation} \label{effective potential gravitational}
V^{(G)} = \frac{P^2}{r^2}\left[\beta(\beta + 1 )-\alpha\frac{r^2}{L^2} \right],
\end{equation} 
where the quantities $\alpha$ and $\beta$ are defined in Eqs.\ (\ref{e:alpha value}) and (\ref{e:beta value}). Using the tortoise coordinate $r_*$  (\ref{e:tortoise-coordinate}) we can show that the effective potential (\ref{effective potential gravitational}) simplifies to
\begin{equation} \label{e:effective potential tortoise}
V^{(G)} = \frac{1}{L^2}\left[\frac{\beta(\beta +1)}{\sinh^2(r_*/L)} - \frac{\alpha}{\cosh^2(r_*/L)}\right].
\end{equation} 

When $D=4$, the de Sitter QN frequencies (\ref{e:quasinormal-gravitational}) and the effective potential (\ref{e:effective potential tortoise}) for a vector type gravitational perturbation reduce to those for an axial gravitational perturbation given in Appendix B of Ref.\ \cite{Lopez-Ortega:2006ig}.

\section{Scalar field}
\label{s: Klein Gordon}

In this section we study the de Sitter QNMs of a coupled to curvature, real massive scalar field. These modes were already studied in Refs.\ \cite{Du:2004jt}, \cite{Abdalla:2002hg}, but here we discuss some additional facts and make some comments on our results and those of the previous references. 

The Klein-Gordon equation is 
\begin{equation} \label{e:Klein-Gordon-equation}
(\square^2 + \mu^2 + \xi \mathcal{R})\, \Phi = 0,
\end{equation} 
where $\mu$ is the particle mass, $\mathcal{R} = D(D-1)/L^2$ is the scalar curvature of the de Sitter spacetime, $\xi$ is the coupling constant between a real Klein-Gordon field and the curvature of the spacetime, $\xi > 0$.

In metric (\ref{e:metric-general}) the Klein-Gordon equation (\ref{e:Klein-Gordon-equation})  takes the form
\begin{equation} \label{csix:Klein-Gordon-expanded}
\left\{\frac{1}{P^2} \partial_t^2 - \frac{1}{r^{D-2}} \partial_r(r^{D-2}P^2 \partial_r) - \frac{1}{r^2} \tilde{\nabla}^2 + m^2 \right\} \Phi = 0,
\end{equation} 
where $\tilde{\nabla}^2$ represent the Laplacian on the $(D-2)$-dimensional sphere, $m^2 = \mu^2 + \xi \mathcal{R}$. Thus the constant $m$ stands for an effective mass. Making the following ansatz for a scalar field \cite{Du:2004jt}, \cite{Abdalla:2002hg}
\begin{equation} \label{e:ansatz-Phi}
\Phi = {\rm e}^{-i\omega t}\,\, Y_{lm}\,\, R(r),
\end{equation} 
we find that the function $R$ must be a solution to the differential equation
\begin{align} \label{e:radial equation KG}
\left\{\left(1 - \frac{r^2}{L^2} \right)\frac{{\rm d}^2}{{\rm d}r^2}  \right. &+ \left(\frac{D-2}{r}-\frac{D r}{L^2} \right)\frac{{\rm d}}{{\rm d}r} \nonumber \\ 
& \left. + \frac{\omega^2}{1-r^2/L^2}  - \frac{l(l+D-3)}{r^2} - m^2 \right\}R=0.
\end{align} 

If in previous equation we make the change of variable $y=r^2/L^2$ (as in Sections \ref{s:electromagnetic-field} and \ref{s:gravitational perturbations}), then this equation is transformed into 
\begin{align} \label{eq:Klein Gordon y}
\left\{y(1-y)\frac{{\rm d}^2}{{\rm d}y^2} \right. &- \frac{1}{2}\left[ (D+1)y - (D-1)\right] \frac{{\rm d}}{{\rm d}y} \nonumber \\ 
& \left.  +\frac{1}{4} \left(\frac{\tilde{\omega}^2}{1-y}-\frac{l(l+D-3)}{y}-\tilde{m}^2 \right) \right\} R = 0,
\end{align}
where $\tilde{m}^2 = m^2 L^2$ and we define the quantity $\tilde{\omega}$ as in the previous sections. Proposing a solution of the form (\ref{e: radial ansatz}) for Eq.\ (\ref{eq:Klein Gordon y}), it is simplified to a hypergeometric type differential equation (\ref{e:hypergeometric-differential}) with parameters $a$, $b$, and $c$ equal to \cite{Du:2004jt}, \cite{Abdalla:2002hg}
\begin{align} \label{e:a-b-c-parameters-Klein-Gordon}
a&= A + B + \frac{D-1}{4} + \frac{1}{2}\left[ \frac{(D-1)^2}{4} - \tilde{m}^2 \right]^{\tfrac{1}{2}}, \nonumber \\ 
b&= A + B + \frac{D-1}{4} - \frac{1}{2}\left[ \frac{(D-1)^2}{4} - \tilde{m}^2 \right]^{\tfrac{1}{2}},\\
c&= 2 A + \frac{D-1}{2},  \nonumber
\end{align}
where
\begin{align}
A &= \left\{ \begin{array}{l}\hspace{.5cm} \frac{l}{2} , \\ \\ -\frac{D-3+l}{2}, \end{array} \right.  & B = \pm \frac{i\tilde{\omega}}{2}.
\end{align}

Therefore, as in the previous sections, we have several options for the values of the parameters $a$, $b$, and $c$ depending on the values chosen for $A$ and $B$. In what follows we study in detail the case $A = \tfrac{l}{2}$ and $B=\tfrac{i\tilde{\omega}}{2}$. For this value of $A$ the quantity $c$ is a positive integer for odd $D$ or a half-integer for even $D$. Thus the first solution to the hypergeometric type differential equation is of the form (\ref{e:first-solution}). The second solution is given by (\ref{e:solution-two-I}) when $c$ is a half-integer and by (\ref{e:solution-two-II}) when $c$ is an integer.

An analysis of the radial functions shows that only the function
\begin{equation} \label{e: regular function}
R(r) = y^{\tfrac{l}{2}} (1-y)^{\tfrac{i \tilde{\omega}}{2}}  {}_2F_1(a,b;c;y),
\end{equation} 
is regular at the origin of de Sitter spacetime. Using the definition of Subsection \ref{s:modes-I}, we now study the QNMs of a real scalar field propagating in $D$-dimensional de Sitter spacetime. Its QN frequencies were calculated in Refs.\ \cite{Du:2004jt}, \cite{Abdalla:2002hg}, but here we discuss them in more detail. 

We first consider a real massless scalar field, not coupled to scalar curvature (thus $\mu = \xi= \tilde{m} = 0 $). The regular radial function $R$ at the origin is (\ref{e: regular function}), and the expressions for the parameters $a$, $b$, and $c$ appear in Eq.\ (\ref{e:a-b-c-parameters-Klein-Gordon}) taking $\tilde{m} = 0$.

From the property of the hypergeometric function (\ref{e:hypergeometric-property-z-1-z}), we can expand the radial function $R$ of Eq.\ (\ref{e: regular function}) as in (\ref{e:physical-field-1-y}). Nevertheless, from Eqs.\ (\ref{e:a-b-c-parameters-Klein-Gordon}) it is possible to show that when the quantity $c-a-b$ is not an integer, again we cannot satisfy the conditions to obtain a purely outgoing wave as $r \to L$ (see Sections \ref{s:electromagnetic-field} and \ref{s:gravitational perturbations}). 

If the quantity $c-a-b$ is a negative integer, following the procedure of Subsection \ref{s:modes-I}, we first write the radial function (\ref{e: regular function}) as in Eq.\ (\ref{e:hypergeometric-integer-I}) and then impose again the conditions (\ref{e:conditions-quasinormal-I}) to fulfill the de Sitter QNMs boundary conditions. Hence we can find that the QN frequencies of a massless and not coupled to curvature real scalar field are 
\begin{equation} \label{e:massless-quasinormal-frequencies}
i \tilde{\omega} = l + 2n + (D-1), \qquad \qquad i \tilde{\omega} = l + 2n.
\end{equation} 
For $D\geq 5$, these frequencies are equal to those of a tensor type gravitational perturbation (\ref{e:quasinormal-gravitational}).

In Eqs.\ (\ref{e:quasinormal-frequency-I}), (\ref{e:quasinormal-frequency-II}), (\ref{e:quasinormal-gravitational}), and (\ref{e:massless-quasinormal-frequencies}) we give two expressions for the de Sitter QN frequencies. Notice that in even dimensions these expressions produce different sets of QN frequencies. In odd dimensions except for some frequencies, whose number depend upon the spacetime dimension, both expressions yield the same de Sitter QN frequencies. As already pointed out in the previous Section, for the gravitational perturbations in odd dimensional de Sitter spacetimes they give the QN frequencies calculated by Nat\'ario and Schiappa in \cite{Natario:2004jd}.

For a coupled to curvature, real massive Klein-Gordon field, the regular radial functions at the origin appear in Eq.\ (\ref{e: regular function}) (when $A=\tfrac{j}{2}$ and $B=\tfrac{i \tilde{\omega}}{2}$), however the parameters $a$, $b$, and $c$ have the values (\ref{e:a-b-c-parameters-Klein-Gordon}) with an effective mass $\tilde{m}$ different from zero. Exploiting the property of the hypergeometric functions (\ref{e:hypergeometric-property-z-1-z}), we can write the radial function $R$ as in Eq.\ (\ref{e:physical-field-1-y}) and therefore to satisfy the boundary condition of the de Sitter QNMs near the cosmological horizon, it suffices that either of the following conditions holds 
\begin{equation} \label{e:Klein Gordon-conditions}
c-a=-n, \qquad \textrm{or} \qquad c-b=-n.
\end{equation} 
Here we can exploit the property (\ref{e:hypergeometric-property-z-1-z}) of the hypergeometric function instead of (\ref{e:hypergeometric-property-integer}), because usually the quantity $\sqrt{\left(\tfrac{D-1}{2}\right)^2 - \tilde{m}^2}$ is not an integer and therefore we can fulfill the conditions (\ref{e:Klein Gordon-conditions}) for non-integral $c-a-b$. There are limits where it is not valid, but in these limits we can study the problem as in the massless and not coupled to curvature case (and as in the previous sections).

From Eqs.\ (\ref{e:Klein Gordon-conditions}) we find that the QN frequencies are
\begin{equation} \label{e:quasinormal-massive}
i\tilde{\omega} =  l + 2 n +\frac{D-1}{2} \pm \left( \frac{(D-1)^2}{4}-\tilde{m}^2\right)^{\tfrac{1}{2}} ,
\end{equation} 
for $\tfrac{(D-1)^2}{4} > \tilde{m}^2$ and
\begin{equation} \label{e:quasinormal-massive-two}
i\tilde{\omega} =  l + 2 n +\frac{D-1}{2} \pm i \left( \tilde{m}^2 - \frac{(D-1)^2}{4} \right)^{\tfrac{1}{2}} ,
\end{equation} 
for $\tilde{m}^2 > \frac{(D-1)^2}{4}$.

The energy momentum tensor for a scalar field is equal to \cite{b:Birrell-1982-book-QFT-CST}
\begin{align} \label{e:tensor Klein Gordon}
T_{\mu \nu} &= (1-2\xi) \nabla_\mu\Phi \nabla_\nu \Phi + (2 \xi - \tfrac{1}{2})g_{\mu \nu} g^{\rho \sigma} \nabla_\rho\Phi \nabla_\sigma \Phi - 2 \xi \nabla_\mu \nabla_\nu \Phi \nonumber \\
& + \frac{2}{D} \xi g_{\mu \nu} \Phi \, \square \Phi -\xi\left[ R_{\mu \nu} - \tfrac{1}{2} \mathcal{R} g_{\mu \nu} + \tfrac{2(D-1)}{D} \xi \mathcal{R} g_{\mu \nu} \right] \Phi^2 \nonumber \\
& + \left[ \tfrac{1}{4} - \left(1 - \tfrac{1}{D}\right) \xi \right] \mu^2 g_{\mu \nu} \Phi^2  .
\end{align}   
Using this, we can show that the radial flux is different from zero for the solutions to the Klein-Gordon equation of the form (\ref{e: regular function}) when the frequency of the field is given by any of expressions (\ref{e:massless-quasinormal-frequencies}), (\ref{e:quasinormal-massive}), or (\ref{e:quasinormal-massive-two}). Therefore these frequencies are quasinormal. 

To find the radial flux of the scalar field, it is more convenient to use the component $T^r_t$ of the energy momentum tensor (\ref{e:tensor Klein Gordon}) instead of the conserved current $J_r=(\Phi^* \partial_r \Phi - \Phi \partial_r \Phi^* )/2i$ of Ref.\ \cite{Du:2004jt}, because this last current is identically zero for purely imaginary frequencies. Thus to obtain a  radial flux different from zero in \cite{Du:2004jt} Du, \textit{et al.\ }must choose the de Sitter QN frequencies with real part different from zero. Therefore they discarded the purely imaginary de Sitter QN frequencies. From the results of  previous sections and Refs.\ \cite{Natario:2004jd}, \cite{Lopez-Ortega:2006ig} we must consider that the QN frequencies can be purely imaginary.

Our results show that there are well defined QN frequencies for all the values of $m$, provided we can consider that the scalar field is propagating in a fixed background. This conclusion is different from that of Ref.\ \cite{Du:2004jt} where it is asserted that the de Sitter QN frequencies are well defined only for $\tilde{m}^2 > \frac{(D-1)^2}{4}$.

For the minimally coupled massive scalar field notice that taking the limit $M \to 0$ of the result obtained for its de Sitter QN frequencies, we get that for the minimally coupled massless Klein-Gordon field. Furthermore, if the conclusion of Ref.\ \cite{Natario:2004jd} on the non-existence of the de Sitter QN frequencies for even $D$ also applies to the massless scalar field, then the limit $M \to 0$ of the expression (\ref{e:quasinormal-massive}) for the massive case is not equal to the result of the massless case.

From expressions (\ref{e:quasinormal-massive}) and (\ref{e:quasinormal-massive-two}) we can obtain some limits. For a real massive scalar field coupled to curvature with $\xi = \tfrac{D-1}{4D}$, from formula (\ref{e:quasinormal-massive-two}) we find that the QN frequencies are equal to 
\begin{equation}
i\tilde{\omega} =  l + 2 n +\frac{D-1}{2} \pm i \tilde{\mu}.
\end{equation} 
These quantities take a similar mathematical form to that of the de Sitter QN frequencies for a massive Dirac field in four dimensions (see Section 4.2 of Ref.\ \cite{Lopez-Ortega:2006ig}). In the massless limit, the previous expression yields
\begin{equation} \label{e:massless special}
i\tilde{\omega} =  l + 2 n +\frac{D-1}{2}.
\end{equation} 
Notice that the quantities $i\tilde{\omega}$ given by Eqs.\ (\ref{e:quasinormal-frequency-I}), (\ref{e:quasinormal-frequency-II}), (\ref{e:quasinormal-gravitational}), and (\ref{e:massless-quasinormal-frequencies}) are integral numbers, while the QN frequencies (\ref{e:massless special}) are integers for odd $D$ and half-integers for even $D$. These facts point out that the value of the coupling constant $\xi = \tfrac{D-1}{4D}$ is ``special'' for the de Sitter QNMs of a Klein-Gordon field.

Finally we study a conformally coupled, massless real scalar field ($\xi=\tfrac{D-2}{4(D-1)}$). For this case, from Eq.\ (\ref{e:quasinormal-massive}), the QN frequencies are equal to
\begin{equation} \label{e:quasinormal conformally}
i\tilde{\omega} =  l + 2 n -1 +\frac{D}{2},  \qquad \,\, \qquad i\tilde{\omega} =  l + 2 n +\frac{D}{2},
\end{equation} 
which are half-integers for odd $D$ and integers for even $D$. 

The QN frequencies $\omega$ (\ref{e:quasinormal-frequency-I}), (\ref{e:quasinormal-frequency-II}), (\ref{e:quasinormal-gravitational}), (\ref{e:massless-quasinormal-frequencies}), (\ref{e:quasinormal-massive}), (\ref{e:massless special}) and (\ref{e:quasinormal conformally}) are purely imaginary as those calculated in Refs.\  \cite{Lopez-Ortega:2006ig}, \cite{Fernando:2003ai}, \cite{Saavedra:2005ug}, \cite{Lopez-Ortega:2005ep}, \cite{Cardoso:2001bb}. It is easy to see that for a purely imaginary frequency the relation $\omega = -\omega^*$ holds.

The QN frequencies (\ref{e:quasinormal-massive-two}) are equal to the complex frequencies (16) of Ref.\ \cite{Du:2004jt}. In the cases explicitly studied in the present work we do not find the frequencies (17) of the previous reference; nevertheless, the imaginary part of the complex frequencies (17) of \cite{Du:2004jt} can be negative or positive depending on the values chosen for $n$, $l$, and $D$; thus the amplitude of the field can increase with time for some of these frequencies. This also happens with the QN frequencies (32) of Ref.\ \cite{Abdalla:2002hg}, whose imaginary part can be positive or negative. For all the de Sitter QN frequencies calculated in this paper the amplitude of the field decreases with time.

In Ref.\ \cite{Myung:2003ki} Myung and Kim asserted that de Sitter QNMs do not exist because using real frequencies they cannot satisfy the boundary conditions. As is well known, the QN frequencies are complex numbers \cite{Kokkotas:1999bd}. Moreover, in Ref.\ \cite{Abdalla:2002hg} Abdalla, \textit{et al.}, included as QN frequencies the complex conjugate of the computed frequencies, however as shown in \cite{Kokkotas:1999bd}, if $\omega$ is a QN frequency, then the quantity $-\omega^*$ is also a QN frequency. Hence for the calculated frequencies we believe that its complex conjugate should not be included in the spectrum of the de Sitter QNMs.

We can easily show that Eq.\ (\ref{e:radial equation KG}) simplifies to the form (\ref{e:Schrodinger-equation-I-II}). In de Sitter background the effective potential is equal to \cite{Du:2004jt}
\begin{equation}
V^{(S)} = \frac{1}{L^2}\left[ \frac{l(l+D-3) + \tfrac{(D-2)(D-4)}{4}}{\sinh^2(r*/L)} + \frac{\tilde{m}^2 -\tfrac{D(D-2)}{4} }{\cosh^2(r*/L)} \right].
\end{equation}

\section{Discussion}
\label{s: Discussion}

In Ref.\ \cite{Natario:2004jd} Nat\'ario and Schiappa asserted that the de Sitter QNMs of a gravitational perturbation are only well defined in odd spacetime dimensions. Our results show that the de Sitter QNMs are well defined in even and odd spacetime dimensions for an electromagnetic field, a minimally coupled massless real scalar field and a gravitational perturbation. We believe that the conclusion of Ref.\ \cite{Natario:2004jd} on the non-existence of QNMs in even dimensional de Sitter spacetime is not valid, (and also that of Refs.\ \cite{Brady:1999wd}, \cite{Choudhury:2003wd} for the minimally coupled massless Klein-Gordon field in four dimensions).

For a massive scalar field propagating in $D$-dimensional de Sitter spacetime, in Ref.\ \cite{Du:2004jt} Du, \textit{et al.\ }deduced that there is an inferior limit for its mass to have well defined QN frequencies. In Section \ref{s: Klein Gordon} we show that there are QNMs for a massive real scalar field for all values of its mass (inclusive in the massless case). It is convenient to say that we use a different expression for its radial flux (it comes from the energy momentum tensor (\ref{e:tensor Klein Gordon})) to that utilized in \cite{Du:2004jt}.

We wish to remark that the calculated de Sitter QN frequencies $\omega$ for the massless fields are purely imaginary while those corresponding to massive fields are complex (see also \cite{Lopez-Ortega:2006ig}). Furthermore, it would be interesting to obtain numerical evidence to prove or disprove our results.

Studying the behavior of classical fields in $D$-dimensional spherically symmetric spacetimes, in Refs.\ \cite{Frolov:2002xf}, \cite{Gibbons:2002pq}-\cite{Crispino:2000jx}, \cite{Abdalla:2002hg} is shown that for several physical problems, we get similar results for all values of $D \geq 4$. For instance, the equations of motion for the electromagnetic and gravitational perturbations can be reduced to Schr\"odinger type equations for all $D \geq 4$. Nevertheless, in \cite{Natario:2004jd} is found that the de Sitter QN frequencies for gravitational perturbations are only well defined in odd dimensions. This result allow us to distinguish between even and odd dimensional de Sitter spacetimes. Thus we believe that our result on the existence of well defined QN frequencies for the massless fields in even and odd dimensional de Sitter backgrounds is similar to those of Refs.\ \cite{Frolov:2002xf}, \cite{Gibbons:2002pq}-\cite{Crispino:2000jx}, \cite{Abdalla:2002hg} in the sense that it holds for all $D \geq 4$. 

Notice that the regular radial functions used here and in Ref.\ \cite{Lopez-Ortega:2006ig} satisfy $R(r=0)=0$. We also point out that near the cosmological horizon the outgoing boundary condition for the QNMs of the SdS black hole simplifies to that of the de Sitter QNMs when the radius of the event horizon goes to zero. Nevertheless, we recall that for the SdS spacetime the point $r=0$ is singular, so we cannot compare at this point the boundary conditions for the QN frequencies of the de Sitter and SdS spacetimes. Thus the boundary conditions for the de Sitter QN frequencies used in this paper are not a limit from those of the SdS black hole.

\section{Acknowledgments}

I thank Dr.\ M.\ A.\ P\'erez Ang\'on for his interest in this paper and also for proofreading the manuscript. I also thank Referees for their comments and suggestions. This work was supported by CONACyT and SNI (M\'exico).

\begin{appendix}

\section{Some mathematical facts}
\label{dsd:appendix-mathematical}

In the present paper we often find the hypergeometric differential equation
\begin{equation} \label{e:hypergeometric-differential}
z(1-z) \frac{{\rm d}^2 f}{{\rm d}z^2} + (c - (a +b + 1)z)\frac{{\rm d} f}{{\rm d}z} - a b f   = 0.
\end{equation} 
As is well known, if $c$ is different from zero or $c$ is not a negative integer, then the first solution to the previous differential equation is the hypergeometric function \cite{b:DE-books}
\begin{equation} \label{e:first-solution}
f_1 = {}_{2}F_{1}(a,b;c;z),
\end{equation} 
if $c$ is not an integer, then the second solution is \cite{b:DE-books}
\begin{equation} \label{e:solution-two-I}
f_2= z^{1-c} {}_{2}F_{1}(a-c+1,b-c+1;2-c;z),
\end{equation} 
and when $c$ is an integer greater than zero, the second solution to Eq.\ (\ref{e:hypergeometric-differential}) is \cite{b:DE-books}
\begin{align} \label{e:solution-two-II}
f_2 &= {}_{2}F_{1}(a,b;c;z) \ln (z) \nonumber \\
& + \frac{(c - 1)!}{\Gamma(a) \Gamma(b)} \sum_{s=1}^{c - 1} (-1)^{s-1} (s - 1)! \frac{\Gamma(a - s) \Gamma(b -s)}{(c -s - 1)!} z^{-s} \nonumber \\
&  + \sum_{s=0}^{\infty} \frac{(a)_s (b)_s}{s! (c)_s} z^s \left[\psi(a + s) + \psi(b + s) - \psi(c + s) - \psi(1 + s) \right. \nonumber \\
  & \left. \hspace{0cm} - \psi(a -  c +1) - \psi(b - c + 1) + \psi(1) + \psi(c -1) \right]. 
\end{align} 
In the previous formula $\Gamma(z)$ denotes the Gamma function, $(z)_0=1$, $(z)_s=z(z+1)\ldots(z+s-1)$ for $s\geq 1$ and $\psi(z)={\rm d}\, \textrm{ln} \Gamma(z)/{\rm d}z$.

If the quantity $c-a-b$ is not an integer, then the relation between two hypergeometric functions with variables $z$ and $1-z$ is \cite{b:DE-books}
\begin{eqnarray} \label{e:hypergeometric-property-z-1-z}
{}_2F_1(a,b;c;z) = \frac{\Gamma(c) \Gamma(c-a-b)}{\Gamma(c-a) \Gamma(c - b)} {}_2 F_1 (a,b;a+b+1-c;1-z) \hspace{1cm} \nonumber \\
+ \frac{\Gamma(c) \Gamma( a +b - c)}{\Gamma(a) \Gamma(b)} (1-z)^{c-a -b} {}_2F_1(c-a, c-b; c + 1 -a-b; 1 -z),
\end{eqnarray}
and when $c-a-b=-n$, $n=1,2,3,\dots$, the following relation holds\footnote{For the expressions valid when $c-a-b$ is equal to zero or a positive integer see Refs.\  \cite{b:DE-books}.} \cite{b:DE-books}
\begin{align} \label{e:hypergeometric-property-integer}
{}_2F_1 (a,b;c;z) &=   \frac{\Gamma(a+b-n) \Gamma(n)}{\Gamma(a)\Gamma(b)} (1-z)^{-n} \sum_{s=0}^{n-1}\frac{(a-n)_s (b-n)_s}{s! (1-n)_s}(1-z)^s  \nonumber \\
& - \frac{(-1)^n \Gamma(a+b-n)}{\Gamma(a-n)\Gamma(b-n)} \sum_{s=0}^\infty \frac{(a)_s (b)_s}{s!(n+s)!}(1-z)^s [\textrm{ln}(1-z)  \nonumber \\
&  -\psi(s+1) -\psi(s+n+1)+\psi(a+s)+\psi(b+s)] .
\end{align}

\section{QNMs of the $D$-dimensional near extreme SdS black hole}
\label{app: near extreme quasinormal}

In Ref.\ \cite{Cardoso:2003sw} (see also \cite{Molina:2003dc}--\cite{Moss:2001ga}) was shown that for a massless scalar field, an electromagnetic field, and an axial gravitational perturbation moving in a four dimensional near extreme SdS black hole, the effective potentials of the Schr\"odinger type equations (\ref{e:Schrodinger-equation-I-II}) take the form
\begin{equation} \label{Poschl-Teller potential}
V = \frac{V_0}{\cosh^2(\kappa_b r_*)},
\end{equation} 
where $\kappa_b$ is the approximate value for the surface gravity of the event horizon. Thus the effective potentials are of  P\"oschl--Teller type \cite{b:Poschl Teller potential}.

For the coupled electromagnetic and gravitational perturbations propagating in a near extreme Reissner-\-Nordstr\"om de Sitter black hole in four dimensions, the effective potentials also reduce to the P\"oschl--Teller form (\ref{Poschl-Teller potential}), as shown in Ref.\ \cite{Molina:2003dc}.

Moreover, in Ref.\ \cite{Molina:2003ff} Molina found that a similar result holds for a massive scalar field moving in $D$-dimensional near extreme SdS background or $D$-dimensional near extreme Reissner-Nordstr\"om de Sitter background. Using this fact, in these references are analytically calculated the QN frequencies of these near extreme black holes taking into account the results of Ferrari and Mashhoon \cite{Mashhoon:1984yo}.

Following \cite{Molina:2003ff}, here we calculate the QN frequencies of an electromagnetic field, a vector type and a tensor type gravitational perturbation propagating in a $D$-dimensional near extreme SdS black hole.\footnote{We notice that in Ref.\ \cite{Molina:2003ff} the symbol $\Lambda$ stands for $3/L^2$, while in Refs.\ \cite{Vanzo:2004fy}, it stands for $(D-1)(D-2)/(2L^2)$.}

The effective potentials for an electromagnetic field moving in a $D$-dimensional near extreme SdS black hole are given in (\ref{e:effective-potentials-I-II}), but now the quantity $P^2$ is equal to
\begin{equation} \label{e: approximate value P}
P^2 = \frac{D-1}{L^2} (r^{ap}_c - r)(r-r^{ap}_b) + O(\delta^3),
\end{equation} 
where $r^{ap}_c$ and $r^{ap}_b$ stand for the approximate values of the cosmological $r_c$ and black hole $r_b$ horizons, respectively, and $\delta$ is a small dimensionless parameter defined by \cite{Molina:2003ff}
\begin{equation*}
\delta=\frac{r_c-r_b}{L}\frac{D-1}{2}.
\end{equation*}
Thus, when the cosmological horizon and the event horizon are very close, the quantity $\delta$ is proportional to the distance between these horizons.

For a vector type and a tensor type gravitational perturbations propagating in a $D$-dimensional near extreme SdS black hole the effective potentials are equal to \cite{Kodama:2003jz}, \cite{Natario:2004jd}
\begin{align} \label{e: gravitational potentials vector}
V^{(V)} &= P^2 \left[\frac{l(l+D-3)}{r^2} + \frac{(D-2)(D-4)}{4}\frac{P^2}{r^2} - \frac{r}{2(D-3)}\frac{{\rm d}^3 P^2}{{\rm d}r^3} \right], \nonumber \\
V^{(T)} &= P^2 \left[\frac{l(l+D-3)}{r^2} + \frac{(D-2)(D-4)}{4}\frac{P^2}{r^2} + \frac{D-2}{2r}\frac{{\rm d}P^2}{{\rm d}r} \right],
\end{align} 
where $P^2$ is given in Eq.\ (\ref{e: approximate value P}).

Using a similar procedure to that of Ref.\ \cite{Molina:2003ff}, in the near extreme limit we can reduce the effective potentials (\ref{e:effective-potentials-I-II}) and (\ref{e: gravitational potentials vector}) to the form (\ref{Poschl-Teller potential}), where
\begin{equation} \label{e: V0 many}
V_0 = \frac{l(l+D-3)\kappa_b(r_c-r_b)}{2 r_b^2},
\end{equation} 
for the modes \textbf{I} of an electromagnetic field, a vector type gravitational perturbation and a tensor type gravitational perturbation, while
\begin{equation} \label{e: V0 one}
V_0 = \frac{(l+1)(l+D-4)\kappa_b(r_c-r_b)}{2 r_b^2},
\end{equation} 
for the modes \textbf{II} of an electromagnetic field.

From the results of \cite{Mashhoon:1984yo}, we find that the QN frequencies for these fields in $D$-dimensional near extreme SdS black hole are equal to
\begin{equation} \label{e: QN Poschl-Teller}
\omega = \kappa_b \left[ \sqrt{\frac{V_0}{\kappa_b^2} - \frac{1}{4}} - i \left( n + \frac{1}{2}\right)\right].
\end{equation} 
In the previous formula $n$ denotes the overtone number.

For a minimally coupled massless scalar field, electromagnetic modes \textbf{I}, vector type and tensor type gravitational perturbations, the QN frequencies in this approximation are equal. Also, note that here we do not calculate the QN frequencies of a scalar type gravitational perturbation owing to the complicated form of the corresponding effective potential.

Taking $D=4$ in (\ref{e: V0 many}) we do not get the result for $V_0$ given in Eq.\ (17) of \cite{Cardoso:2003sw} for a vector type gravitational perturbation moving in a near extremal SdS spacetime (in four dimensions our expression reduces to $V_0=\kappa_b^2 l (l+1)$ while this quantity is equal to $\kappa_b^2 (l+2) (l-1)$ in \cite{Cardoso:2003sw}). The cause of this discrepancy is that here and in Ref.\ \cite{Cardoso:2003sw} are used slightly different approximations to calculate the effective P\"oschl--Teller potentials.

In Ref.\ \cite{Yoshida:2003zz}, Yoshida and Futamase showed that the analytical QN frequencies of a four dimensional near extreme SdS black hole calculated in \cite{Cardoso:2003sw} coincide with those computed numerically when the imaginary part of them is small ($n$ is small), but for large $n$ the analytical and numerical values disagree. Thus from the analytical expression given in Ref.\ \cite{Cardoso:2003sw} we cannot obtain the correct asymptotic behavior ($n \to \infty$) of the QN frequencies. To our knowledge there are no numerical computations of these quantities for the near extreme SdS black hole when $D \geq 5$. Since we use a similar method to that of Ref.\ \cite{Cardoso:2003sw} to calculate the expressions (\ref{e: V0 many}), (\ref{e: V0 one}), and (\ref{e: QN Poschl-Teller}) we believe that the QN frequencies of the $D$-dimensional near extreme SdS black hole calculated here and in \cite{Molina:2003ff} are only valid for small values of the overtone number $n$, and they are inaccurate for large $n$, as in four dimensions.

\section{Effective potentials}
\label{dsd:app-effective-potential}

From the results of Refs.\ \cite{Natario:2004jd}--\cite{Vanzo:2004fy},  \cite{Molina:2003ff}, \cite{Aros:2002te}, \cite{Lopez-Ortega:2006ig}, \cite{Cardoso:2001hn}, \cite{Cardoso:2003sw}, and the present paper, we note that for many spacetimes in which it is possible to find the exact values of the QN frequencies, the Schr\"odinger type radial equation 
\begin{equation} \label{e:Schrodinger-type}
\left[\frac{{\rm d}^2}{{\rm d}x^2} + \mathfrak{a}^2 - V(x) \right] \Phi = 0,
\end{equation} 
has effective potentials of the form
\begin{equation} \label{e:effective-potential}
V(x) = \frac{\mathcal{A}}{\sinh^2(x)} + \frac{\mathcal{B}}{\cosh^2(x)},
\end{equation} 
where the quantities $x$, $\mathfrak{a}^2$, $\mathcal{A}$, and $\mathcal{B}$ appear in Table 1. 

We use in Table 1 the following abbreviations: KG = Klein-Gordon field, EM = electromagnetic field, GP = gravitational perturbation, $M$ is the black hole mass, $q$ is the black hole charge, $Q$ is the eigenvalue of the Laplacian on a manifold of negative constant curvature, $l$ is related to the eigenvalue of the Laplacian on a manifold of positive constant curvature, $\tilde{m}= m L $, $m$ is the effective mass of a Klein-Gordon field (it may include curvature couplings), and $L$ is related to the cosmological constant. The quantities $C_+$ and $C_-$ are equal to \cite{Molina:2003dc}
\begin{equation}
C_{\pm} = (l+2)(l-1)r_b^2 + 3 M r_b \pm r_b \sqrt{9M^2+4(l+2)(l-1)q^2}.
\end{equation} 
We define $C_+$ and $C_-$ to write in a more simple form the value of the parameter $\mathcal{B}$ given in \cite{Molina:2003dc} for the coupled polar modes of the near extreme Reissner-Nordstr\"om de Sitter black hole (see the final rows of Table 1).

We point out that in Ref.\ \cite{Vanzo:2004fy} (\cite{Cardoso:2001hn}) was studied the case of a massless scalar field propagating in Nariai spacetime \cite{b: Nariai solution} (BTZ black hole \cite{Banados:1992wn}); the extension of its results to the case of a massive Klein-Gordon field is straightforward. Moreover, in \cite{Aros:2002te} the effective potential which appears in the Schr\"odinger type equation was not explicitly found, however we can easily calculate it.

Applying the method of Ref.\ \cite{Chandrasekhar book}, we can show in a straightforward way that the equations of motion for a massless Dirac field propagating in four-dimensional and three-dimensional de Sitter spacetime simplify to the form (\ref{e:Schrodinger-equation-I-II}) with the effective potentials equal to \cite{Lopez-Ortega:2006ig}, \cite{Lopez-Ortega:2004cq}
\begin{equation}
V_{\pm} = \frac{\frac{\kappa^2}{L^2}}{\sinh^2(r_*/L)} \mp \frac{\frac{\kappa}{L^2}\cosh(r_*/L)}{\sinh^2(r_*/L)},
\end{equation} 
where $\kappa$ is an integer (half-integer) in four (three) dimensions. In these variables this potential does not take the form (\ref{e:effective-potential}).

Again following the procedure of Ref.\ \cite{Chandrasekhar book} for a massless Dirac field moving in BTZ black hole, we can simplify the equations of motion to the form (\ref{e:Schrodinger-equation-I-II}), with effective potentials equal to \cite{Cardoso:2001hn}
\begin{equation}
V_{\pm} = \frac{\frac{l^2}{L^2}}{\cosh^2\left(\frac{M^{1/2}}{L} r_*\right)} \pm \frac{\left(\frac{lM^{1/2}}{L^2}\right) \sinh\left(\frac{M^{1/2}}{L} r_*\right)}{\cosh^2\left(\frac{M^{1/2}}{L} r_*\right)}.
\end{equation} 
For this case is possible to find exact solutions to the radial differential equations \cite{Cardoso:2001hn}, but an analytical calculation of the QN frequencies was not carried out due to mathematical difficulties, (in the previous reference Cardoso and Lemos calculated numerically the QN frequencies of a massless Dirac field).

For $\mathcal{A}=0$, the potential (\ref{e:effective-potential}) is known in the QNMs literature as P\"oschl--Teller potential \cite{b:Poschl Teller potential}, and it has been used as a first approximation to the effective potentials of several fields propagating in Schwarzschild and SdS black holes to calculate the values of its QN frequencies \cite{Cardoso:2003sw}, \cite{Moss:2001ga}, \cite{Mashhoon:1984yo}, \cite{Zhidenko:2003wq}, (see previous Appendix).

As noticed in Ref.\ \cite{Lopez-Ortega:2006ig}, Beyer showed that for a potential of the form (\ref{e:effective-potential}) with $\mathcal{A}=0$, the QNMs form a complete set \cite{Beyer:1998nu}. If a similar result holds for the potential (\ref{e:effective-potential}) when both $\mathcal{A}$ and $\mathcal{B}$ are different from zero is an interesting question. Finally, are there other spacetimes to include in Table 1?


\begin{landscape}
\begin{longtable}{p{2.9cm} p{1.5cm} p{1.65cm} p{3.85cm} p{6.1cm}} 
\hline\noalign{\smallskip}
Field  & $ x$ & $ \mathfrak{a}^2$ & $ \mathcal{A}$ & $\mathcal{B}$ \\[3pt] \hline   
\multicolumn{5}{c}{{\small \textbf{$D$-dimensional de Sitter spacetime}, \cite{Natario:2004jd}, \cite{Du:2004jt}, \cite{Lopez-Ortega:2006ig}} }\\ \hline \noalign{\smallskip}
Massive KG & $\frac{r_*}{L}$ &$ (\omega L)^2 $ & $l(l+D-3)  + \frac{(D-2)(D-4)}{4}$ & $-\frac{D(D-2)}{4} + \tilde{m}^2$ \\ \hline\noalign{\smallskip}
EM (Modes \textbf{I}) &$\frac{r_*}{L}$ & $ (\omega L)^2 $ & $l(l+D-3) + \frac{(D-2)(D-4)}{4}$ & $- \frac{(D-4)(D-6)}{4}$\\ \hline\noalign{\smallskip}
EM (Modes \textbf{II}) &$\frac{r_*}{L}$  & $ (\omega L)^2 $& $ l(l+D-3) + \frac{(D-2)(D-4)}{4}$ & $- \frac{(D-2)(D-4)}{4}$\\ \hline\noalign{\smallskip}
GP (Scalar) &$\frac{r_*}{L}$ & $ (\omega L)^2 $ &$l(l+D-3) + \frac{(D-2)(D-4)}{4}$ & $- \frac{(D-4)(D-6)}{4}$\\ \hline\noalign{\smallskip}
GP (Tensor) & $\frac{r_*}{L}$&$ (\omega L)^2 $ &$l(l+D-3) + \frac{(D-2)(D-4)}{4}$ &$- \frac{(D)(D-2)}{4}$ \\ \hline\noalign{\smallskip}
GP (Vector) & $\frac{r_*}{L}$& $ (\omega L)^2 $&$l(l+D-3) + \frac{(D-2)(D-4)}{4}$ & $- \frac{(D-2)(D-4)}{4}$\\ \hline\noalign{\smallskip}
\multicolumn{5}{c}{{\small\textbf{$D$-dimensional Nariai spacetime} \cite{Vanzo:2004fy} }}\\ \hline\noalign{\smallskip}
GP (Tensor) & $\frac{\sqrt{D-1}}{L} r_* $& $ (\frac{\omega L}{\sqrt{D-1}})^2 $& $ 0$ & $ \frac{l(l+D-3)}{(D-3)}$\\ \hline\noalign{\smallskip}
Massive KG & $\frac{\sqrt{D-1}}{L} r_* $& $ (\frac{\omega L}{\sqrt{D-1}})^2 $& $ 0$ & $ \frac{l(l+D-3)}{(D-3)} + \frac{\tilde{m}^2}{D-1}$ \\ \hline\noalign{\smallskip}
\multicolumn{5}{c}{{\small\textbf{$D$-dimensional massless topological black hole}, \cite{Aros:2002te}} }\\ \hline\noalign{\smallskip}
Massive KG & $ \frac{r_*}{L}$ & $ (\omega L)^2 $ & $\frac{(D-2)D}{4} +  \tilde{m}^2$ & $Q - \frac{(D-2)(D-4)}{4}$ \\ \hline\noalign{\smallskip}
\multicolumn{5}{c}{{\small\textbf{Static BTZ black hole}, \cite{Cardoso:2001hn}}}\\ \hline\noalign{\smallskip}
Massive KG &$\frac{M^{1/2} r_*}{L}$ & $\left( \frac{ \omega L}{M^{1/2}} \right)^2$ & $\frac{3}{4}+\tilde{m}^2$ & $ \frac{1}{4} + \frac{l^2}{M} $ \\ \hline\noalign{\smallskip}
\multicolumn{5}{c}{{\small\textbf{Near extreme SdS black hole in 4$D$} \cite{Cardoso:2003sw}} }\\ \hline\noalign{\smallskip}
KG and EM & $ r_*$ & $  \omega^2 $ &  $ 0$ & $\kappa_b^2 l (l+1)$ \\ \hline\noalign{\smallskip}
GP (Axial) & $ r_*$ & $ \omega^2 $ & $ 0$ & $\kappa_b^2 (l+2) (l-1)$ \\ \hline\noalign{\smallskip}
\multicolumn{5}{c}{{\small\textbf{$D$-dimensional near extreme SdS black hole} \cite{Molina:2003ff} } }\\ \hline\noalign{\smallskip}
Massive KG & $ r_*$ & $  \omega^2 $ & $ 0$ & $\frac{l(l+D-3)}{r_b^2}\frac{(r_c-r_b)\kappa_b}{2}$  $+m^2\frac{(r_c-r_b)\kappa_b}{2}$ \\ \hline\noalign{\smallskip}
EM (Modes \textbf{I}) & $ r_*$ & $\omega^2 $ & $ 0$ & $\frac{l(l+D-3)}{2r_b^2} (r_c-r_b)\kappa_b$ \\ \hline\noalign{\smallskip}
EM (Modes \textbf{II}) & $ r_*$ & $ \omega^2 $ & $ 0$ & $\frac{(l+1)(l+D-4)}{2r_b^2} (r_c-r_b)\kappa_b$ \\ \hline\noalign{\smallskip}
GP (Vector and Tensor) & $ r_*$ & $ \omega^2 $ & $ 0$ & $\frac{l(l+D-3)}{2r_b^2} (r_c-r_b)\kappa_b$ \\ \hline\noalign{\smallskip}
\multicolumn{5}{c}{{\small\textbf{Near extreme RNdS black hole in $4D$} \cite{Molina:2003dc}} } \\ \hline\noalign{\smallskip}
KG & $ r_*$ & $  \omega^2 $ & $ 0$ & $\frac{l(l+1)}{r_b^2}\frac{(r_c-r_b)\kappa_b}{2} $ \\ \hline\noalign{\smallskip}
Coupled EM and GP (Axial modes 1) & $ r_*$ & $  \omega^2 $ & $ 0$ & $[l(l+1)r_b^2+4q^2 - r_b(3M - \sqrt{9M^2+4(l+2)(l-1)q^2}) ]\frac{(r_c-r_b)\kappa_b}{2r_b^4} $ \\ \hline\noalign{\smallskip}
Coupled EM and GP (Axial modes 2) & $ r_*$ & $  \omega^2 $ & $ 0$ & $[l(l+1)r_b^2+4q^2 -r_b(3M + \sqrt{9M^2+4(l+2)(l-1)q^2}) ]\frac{(r_c-r_b)\kappa_b}{2r_b^4} $ \\ \hline\noalign{\smallskip}
Coupled EM and GP (Polar modes 1) & $ r_*$ & $  \omega^2 $ & $ 0$ & $ \frac{(r_c-r_b)\kappa_b}{2r_b^4} \times \left[(2Mr_b -2q^2-2\Lambda r_b^4/3) \right.$ $ \left. + \frac{(C_+)((l+2)(l-1)r_b/2 + M + 2\Lambda r_b^3/3)}{(l-2)(l+1)r_b/2 + 3M -2q^2/r_b} \right] $ \\ \hline\noalign{\smallskip}
Coupled EM and GP (Polar modes 2) & $ r_*$ & $  \omega^2 $ & $ 0$ &  $ \frac{(r_c-r_b)\kappa_b}{2r_b^4} \times \left[ (2Mr_b -2q^2-2\Lambda r_b^4/3) \right. $ $\left. + \frac{(C_-)((l+2)(l-1)r_b/2 + M + 2\Lambda r_b^3/3)}{(l-2)(l+1)r_b/2 + 3M -2q^2/r_b} \right] $ \\ \hline\noalign{\smallskip}
\caption{Parameters for the Schr\"odinger differential equation (\ref{e:Schrodinger-type}).}
\label{t:table}
\end{longtable}
\end{landscape}

\end{appendix}

\end{document}